\documentclass[10pt,a4paper,final]{iopart}

\usepackage{iopams}  
\usepackage{graphicx}
\usepackage[breaklinks=true,colorlinks=true,linkcolor=blue,urlcolor=blue,citecolor=blue]{hyperref}
\usepackage[margin=0.7in]{geometry}
\usepackage{caption}
\usepackage[font=small]{caption}

\begin{document}

\title{The 3D Kasteleyn transition in dipolar spin ice: a numerical study with the Conserved Monopoles Algorithm.}

\author{M. L. Baez} 
\address{Dahlem Center for Complex Quantum Systems, Freie Universit\"at Berlin, Germany, and Helmholtz-Zentrum f\"ur Materialien und Energie, Berlin, Germany.}

\author{R. A. Borzi}
\address{Instituto de Investigaciones Fisicoqu\'\i{}micas Te\'oricas y 
Aplicadas (INIFTA), UNLP-CONICET and\\
Departamento de F\'\i{}sica, Facultad de Ciencias Exactas, Universidad 
Nacional de La Plata,  c.c.\ 16, suc.\ 4, 1900 La Plata, Argentina. \footnote{Current address: Instituto de F\'{\i}sica de L\'{\i}quidos y Sistemas Biol\'ogicos (IFLYSIB), UNLP-CONICET, La Plata, Argentina}}
\ead{{borzi@fisica.unlp.edu.ar}}

\begin{abstract}
We study the three-dimensional Kasteleyn transition in both nearest neighbours and dipolar spin ice models using an algorithm that conserves the number of excitations. We first limit the interactions range to nearest neighbours to test the method in the presence of a field applied along $[100]$, and then focus on the dipolar spin ice model. The effect of dipolar interactions, which is known to be greatly self screened at zero field, is particularly strong near full polarization. It shifts the Kasteleyn transition to lower temperatures, which decreases $\approx 0.4 K$ for the parameters corresponding to the best known spin ice materials, $Dy_2Ti_2O_7$ and $Ho_2Ti_2O_7$. This shift implies effective dipolar fields as big as $0.05$ tesla opposing the applied field, and thus favouring the creation of ``strings" of reversed spins. We compare the reduction in the transition temperature with results in previous experiments, and study the phenomenon quantitatively using a simple molecular field approach. Finally, we relate the presence of the effective residual field to the appearance of string-ordered phases at low fields and temperatures, and we check numerically that for fields applied along $[100]$ there are only three different stable phases at zero temperature. 

\end{abstract}

\vspace{2pc}
\submitto{\JPCM}
\maketitle

\section{Introduction}

\begin{figure}[htb!]
\centering
\includegraphics[scale=0.3,angle=0]{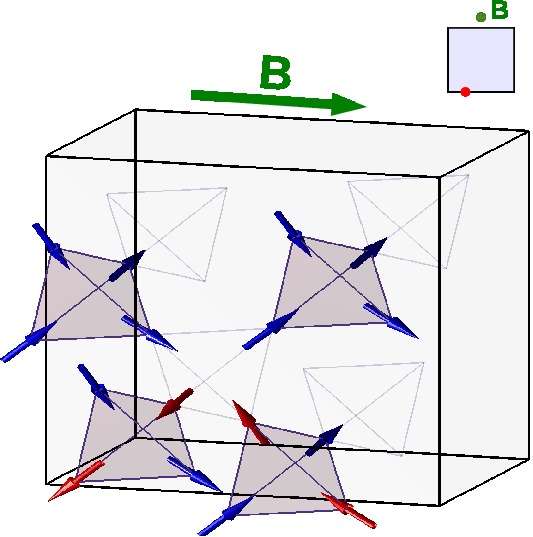}
\caption{The pyrochlore cubic unit cell. Ising spins sit on the shared corners of tetrahedra running along the local $\langle 111 \rangle$ directions. There are two types of tetrahedron, each type conforming an fcc sublattice. The spin configuration shown corresponds to the half polarized phase (see text) that can be stabilised with a moderate field {\bf B} applied along $[100]$ at low temperature. It can be understood by starting with a fully polarized state along the $[100]$ direction  (blue arrows), and then flipping a helical string  of spins with axis along $[100]$ (red arrows). The top right shows a 2D sketch of this configuration as seen from the field direction: we represent the inverted string on the otherwise empty unit cell by a red dot.}
\label{semipol}.
\end{figure}

Spin ice materials belong to the family of geometrically frustrated magnetic systems \cite{diep}. $Dy_2Ti_2O_7$ and $Ho_2Ti_2O_7$ were first signaled out for their measured residual magnetic entropy, similar to that calculated by Pauling for water ice \cite{Ramirez}.  Their magnetic behaviour below $\approx 10 K$ can be described by Ising-like magnetic moments or \emph{spins} arranged in a pyrochlore lattice, the cubic unit cell of which we show in Fig.~\ref{semipol}. The spin quantization axis points along the local $\langle 111 \rangle$ directions so that each spin can point either into or out of the tetrahedron in which vertex sits \cite{bram_ging}. Considering only nearest neighbours ferromagnetic interactions between spins ---restriction that defines the \emph{nearest neighbours} (NN) spin ice  model (NNSIM)--- results in the stabilization of an exponentially degenerate ground state. Its associated manifold of states consists of spin configurations in which two spins point into and two out of each tetrahedron ---the \emph{spin ice} (SI) \emph{rule}. Within the NNSIM, the only excitations at zero magnetic field correspond to breaking the local balance between inward and outward spins. A single positive (negative) excitation or \emph{monopole} consists of three spins pointing into (out of) a tetrahedron, while the fourth spin points in the opposite direction. While double charge excitations are also possible, we will not need to consider them for the purpose of this work.

Even though the SI rule is not enough to impose an ordered ground state even at the lowest temperatures, it does establish a strong degree of spin correlation, which mimics that in magnetic systems near criticallity \cite{bowtie2,bowtie3}. While space correlations generally require subtle probes to be noticed \cite{bowtie}, the marked contrast between spin ice and a simple paramagnet was beautifully made evident by considering the combined effect of applying a field ($B$) along the $[100]$ direction and imposing the SI rule on the NNSIM. Using computer simulations, Jaubert \emph{et al}. showed that in these conditions the magnetisation saturates abruptly in the thermodynamic limit \cite{jaubert}, at a \emph{finite} value of the combined variable field over temperature (\emph{$B/T$}). In this way, it conformed a three dimensional example of a \emph{Kasteleyn transition}, a type of phase change first introduced by Kasteleyn for studying the statistics of a system of non-overlapping dimers on a lattice \cite{kast}. The critical temperature $T_k^{NN}(B)$, below which the system stays in a fully polarized (FP) phase (i.e., with no fluctuations) was shown to be

\begin{equation}
 k_B T_k^{NN} = \frac{2 \mu_{100} B}{log(2)}\,,
\label{Kastel}
\end{equation}

\noindent where $\mu_{100} = \mu / \sqrt{3}$ is the spin's magnetic moment component along the $[100]$ direction, with $\mu$ near $10 \mu_B$ (10 Bohr magnetons) in real spin ices, and $k_B$ is the Boltzmann's constant. Indeed, with monopole point-like excitations banned by the SI rule the only possible way to excite the polarized state is reversing a whole chain of spins against the field direction or creating a \emph{string} (like the one formed by red arrows in Fig.~\ref{semipol}). The temperature $T_k^{NN}$ marks the balance between the two energy scales (the Zeeman effect and the entropic contribution to the free energy) involved in the creation of such one-dimensional excitation of indefinite shape. Above $T_k^{NN}(B)$ ---or below a critical field $B_K^{NN}(T)$, if we invert Eq.~\ref{Kastel}--- the density of these excitations stops being \emph{strictly} zero. The system is then spanned by a macroscopic number of percolating, closed (thanks to periodic boundary conditions) strings.  While there exists some experimental work on this three-dimentional Kasteleyn transition on the spin ice material $Dy_2Ti_2O_7$ ~\cite{CasNature2}, the study is necessarily restricted to relatively high temperatures (where the condition dealing with the imposition of the SI rule is not totally enforced) due to the low temperature freezing of the dynamics (see below).

More realistic numerical simulations of SI materials requires taking into account long range dipolar interactions by using the Ewald summation method \cite{Hertog}. Their inclusion defines the \emph{dipolar spin ice model} (DSIM)  \cite{reviewMelko}.   An important feature appearing on the DSIM is the manifestation of the charge-like character of the localised excitations, now (to a good degree of approximation) \emph{Coulomb interacting} magnetic charges \cite{CasNature2}. 
Approximating the spins by monopole dumbbells of opposite sign, sitting in the centres of neighbouring tetrahedra, defines the \emph{monopolar spin ice model}, MSIM. Within this model, the spin configuration energy is totally determined by the number and relative position of the magnetic charges  \cite{CasNature2}. Quite relevant to the present study is that within the MSIM all configurations satisfying the SI rule have the same magnetic enthalpy. A second feature related to the inclusion of the dipolar forces is the freezing out of the spin dynamics at temperatures below $\approx$ 0.6 K \cite{Mat1,Mat2,Sny}, first studied withing the MSIM  by Jaubert et al. \cite{Jaub-Hold}. Equilibrium at very low temperatures can still be attained at zero applied field using a multiple spin-flip \emph{loop} algorithm \cite{melko,Bakerma} or the \emph{conserved monopoles algorithm}, CMA ~\cite{borzi}. As first shown by Melko and collaborators ~\cite{melko}, dipolar interactions do break the degeneracy of the ground state of the NNSIM, selecting twelve symmetry-connected low energy states (the \emph{MDG phase}) out of the exponentially degenerated manifold \footnote{Another route to equilibrium, so far explored only within the NNSIM both with and without a magnetic field, is through the calculation of the density of states of the system \cite{ferreyra}.}.

Establishing the nature of the lowest energy state for a small $[100]$ field in the DSIM ---the half polarized (HP) phase shown in Fig.~\ref{semipol}, intermediate between the FP and the MDG phase--- required a more sophisticated algorithm, developed only recently \cite{taiwan}. Though not yet explored, it is expected that the long range dipolar interactions (a third energy scale in the system, not included in Eq.~\ref{Kastel}) would also affect the Kasteleyn transition at higher fields. How it alters the simple $T-B$ phase diagram implied by Eq.~\ref{Kastel} is one of the questions we address in the present work. Our approach will also show that the very simple CMA can be implemented to achieve this goal. The method was first introduced to study SI-like systems dominated by a high degree of monopole correlations, with no magnetic field applied \cite{borzi}. It has also been used recently to study artificial square spin ice \cite{xie} and to investigate possible ground states in spin ice materials \cite{borzi1} in the same zero field conditions. We now show that the method can easily be extended to cases with an applied field. 

The manuscript is structured as follows. After introducing the system and methods (Section \ref{sec:methods}) we study the three dimensional Kasteleyn transition in the NNSIM. The explicit contrast of the critical exponents obtained using Monte Carlo simulations with those calculated analytically \cite{jaubert,Har} is used to test the CMA with an applied magnetic field (Section \ref{sec:NNSI}). We then show how dipolar interactions affect the phase diagram, and introduce a new version of Eq.~\ref{Kastel} to describe the Kasteleyn temperature modified by dipolar interactions, modeled as a molecular field (Section \ref{sec:DSI}). Finally, we concentrate on the ground state of the system when a $[100]$ field is applied, checking for states with intermediate magnetisation values other than the HP phase.

\section{Systems and methods}\label{sec:methods}

\subsection{The models}

The DSIM includes dipolar and first neighbours exchange interactions, as well as a Zeeman magnetic energy term proportional to the field $\bf{B}$. We define it by the Hamiltonian:

\begin{equation}
\mathcal{H}_{d}=-J\sum_{<i,j>}\boldsymbol{S}_i\boldsymbol{S}_j+Dr_{nn}^3\sum_{i>j}\frac{\boldsymbol{S}_i\cdot\boldsymbol{S}_j}{|\boldsymbol{r}_{ij}|^3}-\frac{3(\boldsymbol{S}_i\cdot\boldsymbol{r}_{ij})(\boldsymbol{S}_j\cdot\boldsymbol{r}_{ij})}{|\boldsymbol{r}_{ij}|^5} -  \mu \boldsymbol{B} \cdot \sum_{i} \boldsymbol{S_i}\,,
\label{moddip}
\end{equation}

\noindent where $\boldsymbol{S}_i$ are Ising spins with $\langle 111 \rangle$ anisotropy in a pyrochlore lattice (Fig.~\ref{semipol}), $J$ and $D$ are the exchange and dipolar couplings, respectively, $r_{ij}$ is the distance between $\boldsymbol{S}_i$ and $\boldsymbol{S}_j$ spins, and $r_{nn}$ is the nearest neighbours spin distance \cite{Hertog}. If we consider only the nearest neighbours dipolar interactions we derive the effective NNSIM (in the particular case of a field applied along $[100]$):

\begin{equation}
\mathcal{H}_{nn}=J_{eff}\sum_{<i,j>}\sigma_i\sigma_j -  B \mu_{100} \sum_{i} \sigma'_i\,,
\end{equation}

\noindent where $J_{eff}=J_{nn}+D_{nn}=J/3+5D/3>0$ is the effective antiferromagnetic nearest neighbours coupling, $\sigma=\pm1$ is a pseudospin that takes the value $\sigma=1 (-1)$ if the spin point outwards (inwards) of the up tetrahedron were it conventionally belongs. Since we will be working with a constant number of monopoles (see next section) the exchange term will constitute just an energy shift. This implies that only the Zeeman contribution is relevant for the NNSIM (like in a paramagnetic system), while in the DSIM the dipolar part will impose another energy scale.

\subsection{Numerical Method}

In order to limit the exploration of the configuration space of the system to a quasi spin ice rule manifold (in practice, configurations where all but \emph{two} tetrahedron satisfy the spin ice rule) we used the CMA in the limit of zero density of monopoles. Starting from a random configuration that satisfies the ice rule, we flip a spin in order to create just two single monopoles of opposite sign. From then on we follow the usual single spin flip Metropolis algorithm with one extra constraint: we forbid spin flips which either create or destroy more excitations.  The transition between microstates can then be thought as the \emph{effective} movement of the two monopoles around the diamond lattice conformed by the centres of the tetrahedra. 

By construction, the two monopoles of opposite sign start their effective movement from neighbouring tetrahedra. With zero field applied, their random wanderings along the diamond lattice they inhabit describe a single curve, built by joining the neighbouring visited sites. Occasionally, the monopoles would meet again in two neighbouring tetrahedra, closing the open curve into a loop. In an effective way, they would be performing the loop algorithm \cite{melko} in a step by step manner. If a field is applied along [100] the tendency of spins to polarise will be reflected on the trend of the monopole with a plus charge to move along the [100] direction (in analogy to a plus regular charge pushed by an electric field) while the minus one follows (on average) the opposite path. Periodic boundary conditions make this oriented string to wind up while polarizing the system, until full polarization is reached when $B/T$ exceeds a critical threshold. The inclusion of dipolar interactions adds, at first order of approximation (i.e., within the MSIM) a certain attraction for monopoles of opposite sign, without changing this basic picture. Though the method still comprises single spin flips, it does not suffer from dynamical freezing \cite{borzi} and can be used without further modifications (as we will see in this work) in the presence of applied fields.
 
 Within the NNSIM, typical simulation times for a relatively big lattice size (linear latice size $L = 20$) take $\approx 10^5$ Monte Carlo steps for equilibration and a similar number of steps for averaging for each measured point, with further average over 20 independent runs. Note that the restriction to two wandering monopoles is irrespective of the system size. The inclusion of Ewald summations increased markedly the simulation time for the DSIM, forcing us to restrict $L$ to relatively small sizes. Typical times in this case ($L = 5$) take $\approx 10^6$ equilibration Monte Carlo steps and the same for averaging at each measured point, with further averaging over 20 independent runs.
 
We kept units meaningful by using the internal parameters of  $Dy_2Ti_2O_7$, one of the most studied spin ices. We took $D=1.41\,\mathrm{K}$, $r_{nn} =3.54 \AA{}$, $\mu = 10 \mu_B$, and $a=10.01\AA{}$. As mentioned in the previous paragraph, the value of $J$ is unimportant.


\section{Results and discussion}
\label{sec:results}

\subsection{A test for the Conserved Monopoles Algorithm with applied field: the Kasteleyn transition for the nearest neighbours model.}

\begin{figure}[htb!]
\centering
 \includegraphics[scale=0.3]{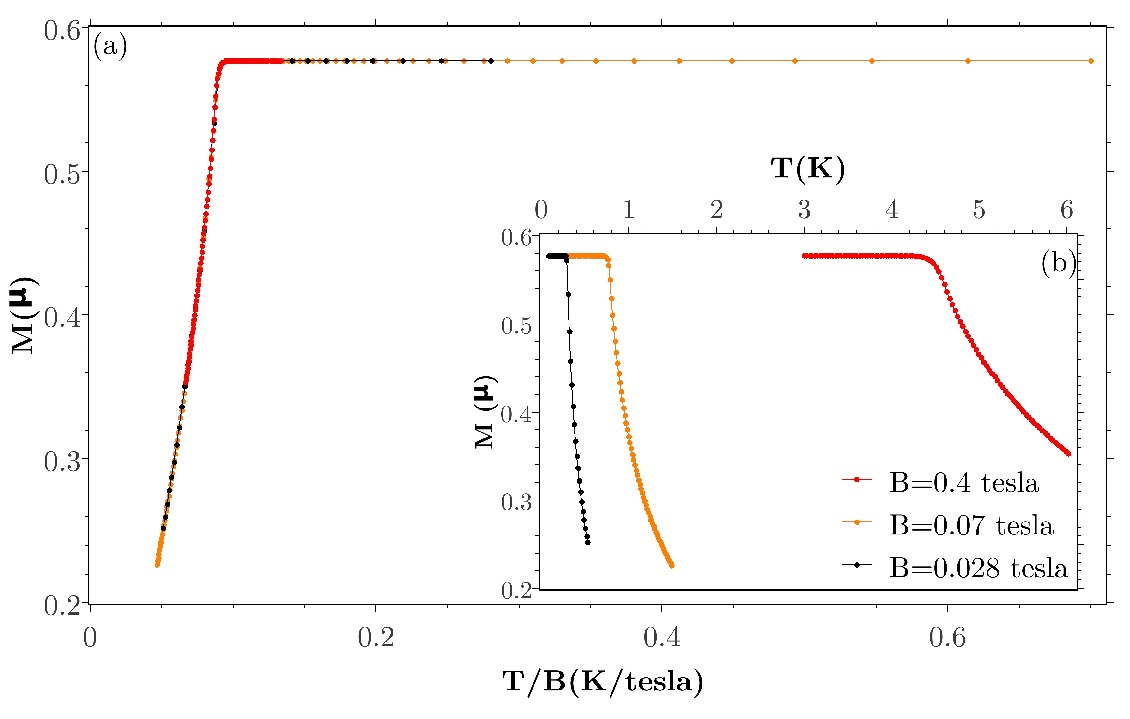}
 \caption{\label{figure1} Magnetisation curves at three different applied fields ($B = 0.028,$ $0.07,$ and $0.4$ tesla) for the NNSI model near the Kasteleyn transition. The lattice size is $L = 20$. (a) Magnetisation as a function of temperature over field. We observe a perfect collapse of the curves indicating the presence of only one energy scale. (b) Magnetization against temperature.}
\end{figure}

As a first approach we use the CMA in the limit of negligible density of monopoles (i.e., within a quasi-perfect spin ice manifold) to study the nearest neighbour model with $\bf{B} // [100]$. The inset to Fig.~\ref{figure1} shows the magnetisation as a function of temperature for different applied fields and size $L = 20$. The tendency to saturate at finite temperatures, with suppressed fluctuations below a critical temperature $T_k^{NN}(B)$, is evident \cite{jaubert}. The transition seems to sharpen for lower applied fields (i.e., lower $T_k^{NN}$, see eq.~\ref{Kastel}). The main figure, where the temperature has been scaled by the magnetic field, implies that this sharpening is but a consequence of the existence of a single energy scale aside from the thermal one. The excellent overlap of the curves shows that the disruption caused by the addition of two monopoles (implying an extra energy scale) is minimal.


\label{sec:NNSI}


\begin{figure}[htb!]
\centering
 \includegraphics[scale=0.3]{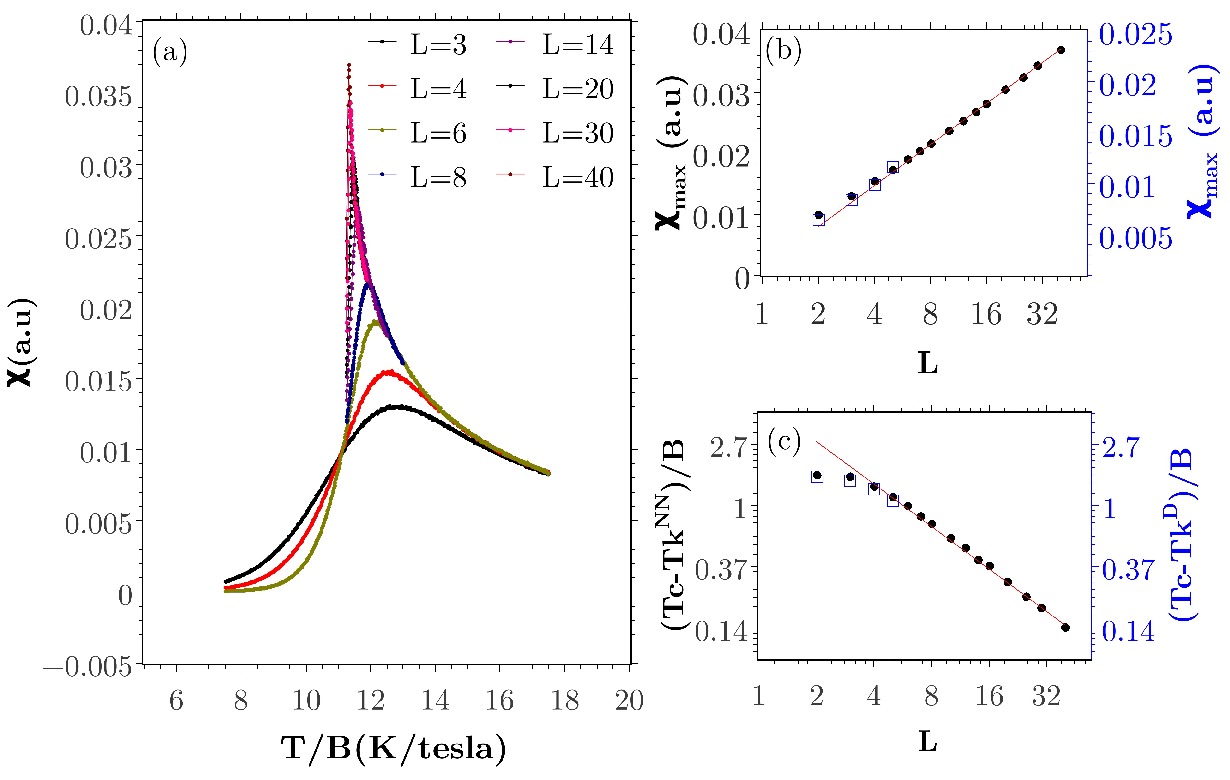}
   \caption{\label{figure2} Finite size effects for the magnetic susceptibility and transition temperature within the NNSIM  ($B = 0.4$ tesla). (a) Magnetic susceptibility as a function of temperature over field and lattice size $L$. Finite size effects are clearly visible, while a big asymmetry between the low and high $T/B$ side develops on increasing $L$. (b) Semi-log plot of the maximum of the susceptibility against L. The red line corresponds to the logarithmic divergence expected for the critical exponent $\gamma = 0$. (c)  Double logarithmic plot of the temperature difference $(T_c-T_k^{NN})/B$ as a function of L. $T_k^{NN}$ is the theoretical critical temperature, and $T_c(L)$ is the temperature at which $\chi$ is maximum. The red line corresponds to $T_c-T_k^{NN} \propto L^{-1/\nu}$, with $\nu = 1$. Both (b) and (c) include data corresponding to the DSIM (open blue squares, right y-axis) measured at $B = 0.62$ tesla. In this field regime, the DSIM shows the same behaviour observed for the NNSIM. The transition temperature expected for the infinite system, modified by the presence of dipolar interactions ($T_k^D$) was not fitted, but imposed using the measured magnetic enthalpy and extrapolating $\Delta/ ln(2)$ (see Eq. (5)) for a macroscopic size.}
\end{figure}
   
\begin{figure}[htb!]
\centering
  \includegraphics[scale=0.3]{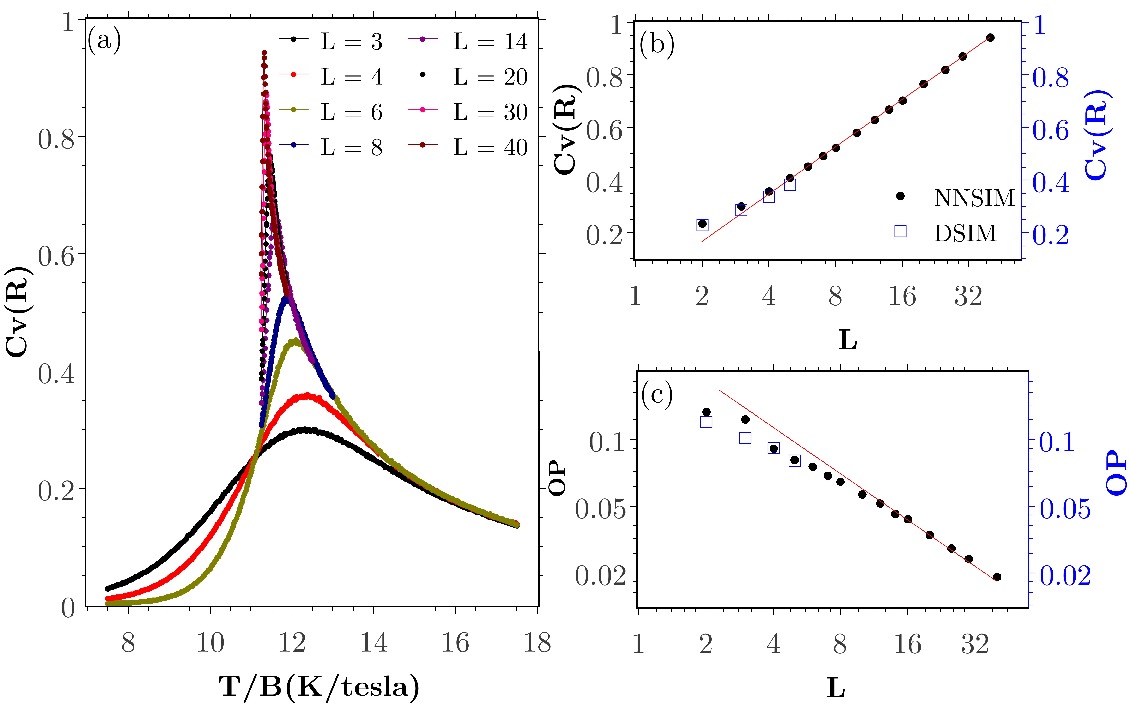}
  \caption{\label{figure2b} Finite size effects for the specific heat and order parameter in the Kasteleyn transition, within the NNSIM. (a) Specific heat as a function of temperature over field, for a magnetic field $B=0.4$ tesla and different lattice size $L$. (b) Semi-log plot of the specific heat evaluated at $T_k^{NN}$ as a function of $L$. The red straight line describes the behaviour expected for a logarithmic divergence ($\alpha = 0$). (c) Log-log plot of the order parameter evaluated at the maximum of the susceptibility as a function of $L$. As a reference, the red line assumes a ratio of critical exponents $\beta/\nu = 1$. Both (b) and (c) include data corresponding to the DSIM (open blue squares, right y-axis), for $B = 0.62$ tesla, showing the same behaviour observed for the NNSIM.}

\end{figure}

As a test for our CMA, we performed a finite size scaling analysis to check compatibility with the known critical temperature and exponents. Fig.~\ref{figure2}a shows the magnetic susceptibility $\chi$ ---the fluctuations of the magnetisation over $T$--- as a function of temperature over field, showing a peak at $T_c(L)$. On increasing system size the peak develops a sharp asymmetry at both sides of the maximum, characteristic of the Kasteleyn transition \cite{ferreyra,jaubert1}. Finite size scaling of the peak height in $\chi$ (Fig.~\ref{figure2}) and specific heat $C_v$ (Figs.~\ref{figure2b}(a) and (b)) allowed us to contrast the critical exponents $\alpha$ and $\gamma$ associated with these quantities for this second order phase transition with those found previously \cite{jaubert,Har,kast-moe,jaubert1}. In both cases the behaviour for big $L$ is compatible with a logarithmic divergence ($\alpha = \gamma = 0$), in coincidence with previous analytical results \cite{jaubert}). Using eq. \ref{Kastel} together with the expression $T_c-T_k^{NN} \propto L^{-1/\nu}$ \cite{binder} we determined a critical exponent compatible with $\nu=1$. We then performed the same analysis for the order parameter, defined as the difference between the saturation and measured magnetisation. Assuming the size dependence scaling as $L^{-\beta/\nu}$, with the value of $\nu$ previously determined, we confirmed \cite{jaubert,jaubert1} the theoretical value $\beta = 1$ (Fig.~\ref{figure2b}(c)). Finally, it should be remembered that the nature of the transition implies an anisotropy between the direction of the field and that perpendicular to it. Identifying the exponent $\nu$ we found (equal to $1$) with $\nu_{\parallel}$ \cite{Har,kast-moe}, we can also determine the correlation length exponent perpendicular to the field ($\nu_{\perp}$) using the scaling law \cite{binder}:

\begin{equation}
\nu_{\parallel}+(d-1)\nu_{\bot}=\gamma+2\beta \rightarrow \nu_{\bot}=\frac{\gamma+2\beta-\nu_{\parallel}}{d-1}\,,
\end{equation}

\noindent and obtain $\nu_{\bot}=1/2$ (\cite{Har,kast-moe}).

The coincidence between these results and those found before gives us confidence on the conserved monopoles method with an applied field. In the following sections we will show that without implying any extra difficulties this algorithm can be used to study how the dipolar interactions affect the behaviour of the Kasteleyn transition, and we will get an insight into the different phases present below $0.3\,\mathrm{K}$.


\subsection{Effect of the dipolar interactions on the Kasteleyn transition.}
\label{sec:DSI}

\begin{figure}[h!]
\centering
 \includegraphics[scale=0.3]{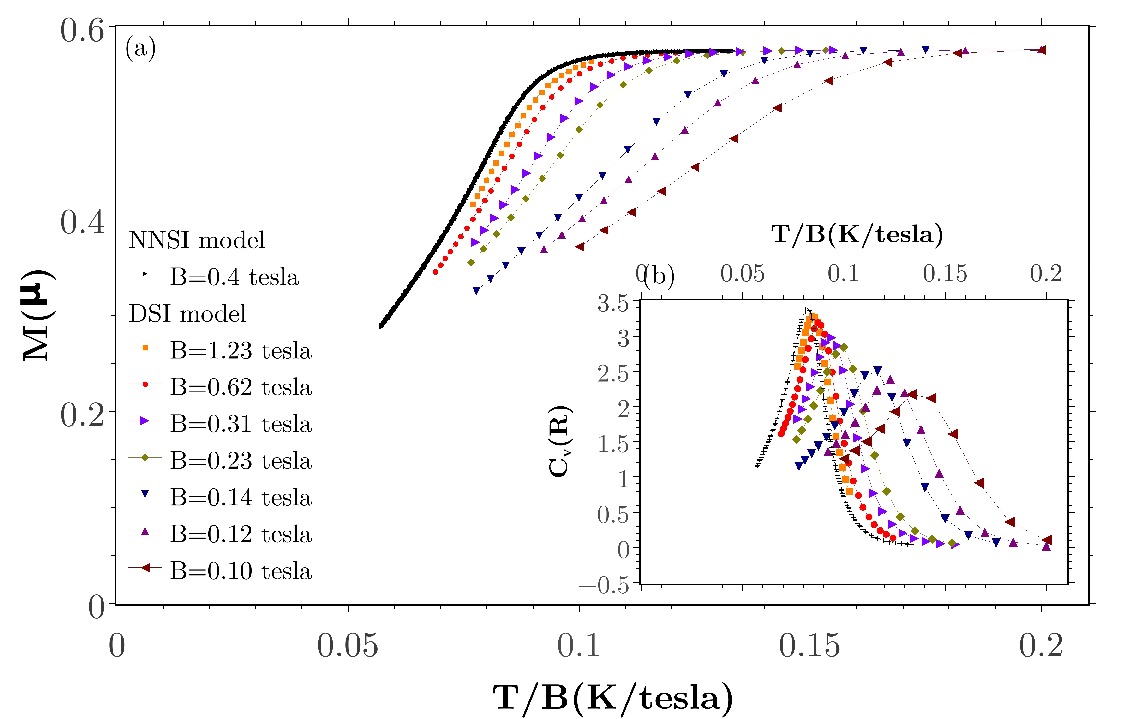}
 \caption{\label{figure3} Magnetisation (a) and specific heat (b) for the DSIM and NNSIM as a function of temperature over field, for different applied fields. The data was taken for a system size $L= 5$. While at high fields both models tend to coincide, there are noticeable differences at low fields, evidencing the relevance of a new energy scale (the dipolar one) in the system.}
\end{figure}

Starting now with the DSIM, Fig.~\ref{figure3}(a) and (b) show the magnetisation and the specific heat as a function of $B/T$, taken at different magnetic fields $\bf{B}$. The NNSIM (single) curve is also included as a reference. It is evident that both models tend to coincide at high fields, where the Kasteleyn transition (which reflects the competition between Zeeman and thermal energies) is expected to take place. On the other hand, there are evident differences at low fields, reflecting the relevance of a new energy scale (the long ranged dipolar interactions) in the system. This is an indication that physics beyond the MSIM is important in this region.

\begin{figure}[h!]
\centering
 \includegraphics[scale=0.3, angle=0]{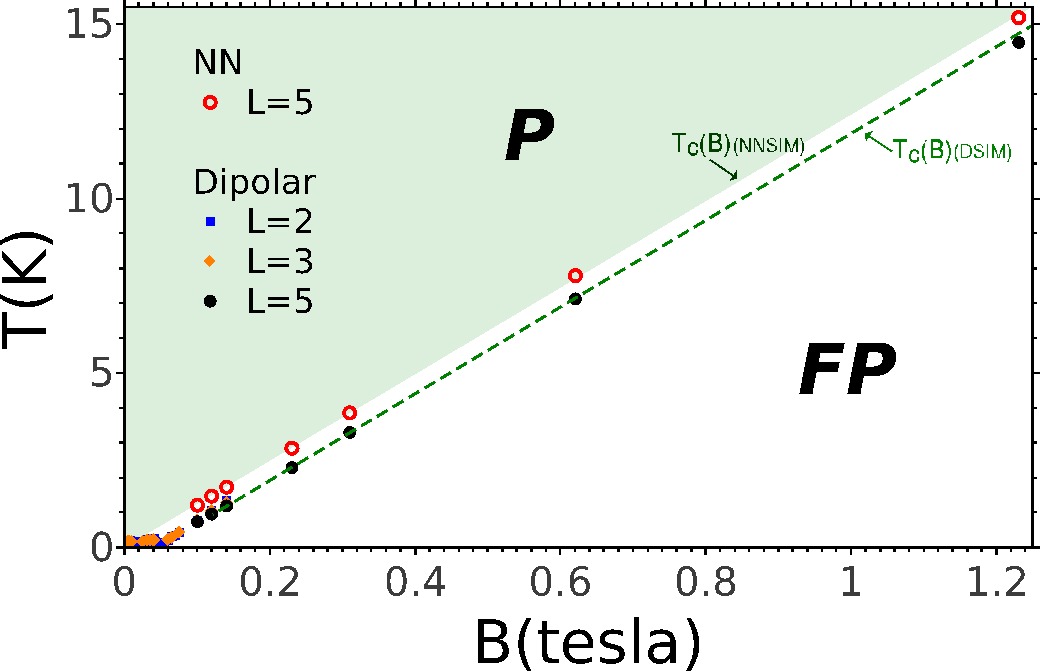}
 \caption{\label{figure5} Phase diagram for the nearest neighbours (NNSIM) and dipolar (DSIM) models, built from the analysis of the specific heat for sizes $L = 2, 3$, and $5$. We have coloured differently the regions corresponding to the paramagnetic (P) and fully polarized (FP) phases for the NNSIM ($L = 5$), limited by the straight line $T_c(B)$ (open circles). In the high field and high temperature regime the DSIM critical temperature also follows a straight line (marked by filled circles and joined by a dashed line), but displaced to lower temperatures ($\Delta T_c \approx -0.46$ K). The low temperature and field region shows some structure corresponding to two other phases, discussed later.}
\end{figure}

Interpreting the maxima of the specific heat curves in (Fig.~\ref{figure3}b) as indications of phase transitions, we have constructed a phase diagram (Fig.~\ref{figure5}). We incorporated data for system sizes $L = 2$, $3$, and $5$ for the DSIM, painting in different colours ---as a reference--- the regions associated with the FP and paramagnetic phase for the NNSIM ($L=5$).

We will concentrate afterwards on the low $T$ and $B$ region, where evidence of the MDG and HP phases (studied on refs. \cite{melko} and \cite{taiwan}) can be observed in Fig.~\ref{figure5}. Following Fig.~\ref{figure3} we expect to recover the Kasteleyn transition for the DSIM at high fields and temperatures. The finite size analysis performed up to $L=5$ (see Figs. \ref{figure2}(b) and (c) and \ref{figure2b}(b) and (c), open blue symbols) shows that this is indeed the case.
Furthermore, we observe in this regime ($B > 0.1$ tesla, $T>0.5\,\mathrm{K}$) a quasi-rectilinear phase boundary. Differently from the NNSIM (open circles), this line does not extrapolate to $T = 0, B = 0$, but compared to Eq.~\ref{Kastel} appears displaced in $\Delta T_c \approx -0.46 ~\mathrm{K}$ (i.e. towards \emph{lower} temperatures). Dipolar interactions thus seem to favour the creation, out of the FP phase, of spin strings anti-parallel to the applied field.  The effect could be understood thinking in an effective magnetic field \emph{reduced} respect to the value of the applied one, an effect that is not surprising given the nature of the dipolar energy (see Eq.~\ref{moddip}). Since the effective field will be useful afterwards to understand better the low temperatures phases, we focus now on this effect.

We will assume that, due to the dipolar interactions,  creating a string out of the FP state leads to a \emph{uniform} magnetic enthalpy gain \emph{per spin in the string} of -$\Delta < 0$. Following ref.~\cite{jaubert}  we will add to this the entropic and Zeeman contribution to the change of free energy associated with the creation of this excitation, in order to find the new (mean field) transition temperature $T_k$:

\begin{equation}
\label{TkDip}
\Delta G_{string}= \Delta U - T_k \Delta S = L(2 \mu_{100}B-\Delta - T_k \ln 2)=L \ln 2 (T_k^{NN}-T_k-\Delta/\ln 2) = 0\,.
\end{equation}

\noindent This mean field treatment leads to $T_k=T_k^{NN}-\Delta/\ln 2$. The predicted constant shift in the transition temperature ($\Delta T_k=-\Delta/\ln 2$) is indeed observed in our data (Fig.~\ref{figure5}).

The inset to Fig.~\ref{figure4} shows the total enthalpy per spin ($U + M B$) for different values of applied field and $L = 5$ as a function of the magnetisation. The focus is near saturation (i.e., near the Kasteleyn transition). While, as expected, the interactions between the two monopoles affect somewhat the magnetic enthalpy for small systems (not shown) this effect is negligible for $L=5$. Curves for different fields $B$ fall on top of each other in Fig.~\ref{figure4}(b), reaching saturation with a linear dependence. It is interesting to note that within the NNSIM, and even taking into account dipolar interactions using the approximate monopolar picture, we would have obtained flat enthalpy curves \cite{CasNature}: the process of string creation takes place at nominally zero monopole density. We conclude then that the observed enthalpy decrease is due to \emph{residual} dipolar interactions (those which cannot be taken into account by an effective Coulomb interaction between monopoles). 
Remarkably, the corrections to the Coulomb potential (which decay as $1/r^5$ \cite{CasNature}) can lead to effects as big as the observed shift in transition temperature ($\approx -0.46 ~K$) or in the transition field ($\Delta B \approx 0.05$ tesla) near full saturation.  

Given the magnitude of the effect of residual dipolar interactions, it is tempting to contrast these results with measurements in real materials. Care should be taken since low temperature experiments would be affected by dynamical arrest, while higher temperatures imply the presence of a finite density of monopolar excitations. It is then remarkable that, within the intermediate temperature window ($T \approx 0.7~K - 1 ~K$),  the study by Morris et al.~\cite{CasNature2} on $Dy_2Ti_2O_7$  also shows a significant shift of $T_k(B)$ towards temperatures lower than the expected $T_k^{NN}(B)$. The size of this shift is about twice of the magnitude we observe \footnote{For example, Fig. 3 in ref.~\cite{CasNature2} implies a saturation field near $0.2$ tesla for $T = 1 ~K$.}. This big discrepancy is most likely due to demagnetisation effects (see Fig. 6.6 in ref. \cite{Demian}). If we use the same criteria employed in ref.~\cite{CasNature2} to determine the saturation field, the subtraction of the demagnetizing effects performed in ref. ~\cite{Demian} gives an estimation of the critical field at $1K$ quite close to the one that can be extracted from Fig.~\ref{figure6}. In spite of this, it is important to stress that a rigorous contrast is difficult, due to possible experimental errors in aligning the applied field and/or the determination of the saturation field for the curves in refs.~\cite{CasNature2,Demian}.

Returning to our simulations, we can go a bit further and obtain an independent estimate of $\Delta$ (and then of $\Delta T_k$)  using the enthalpy shown in the inset of Fig.~\ref{figure4}. We fitted the initial linear decrease in enthalpy from the FP phase (which, due to their dilution, corresponds to the creation of almost independent strings) by a straight line. We thus obtain the enthalpy gain per spin per unit magnetisation along $[100]$. Knowing the magnetisation of a single string to be $-4 L \mu_{100}$, the total number of spins ($16 L^3$), and the number of spins in a string ($4 L$), we can estimate the enthalpy gain \emph{per spin in the string} $\Delta$. For a lattice size $L = 5$ we find $\Delta / ln(2) \approx 0.43 ~K$ -- not far from the value measured in our simulations (Fig.~\ref{figure5}). While there is a certain error in this approach given by finite size effects (we obtain $\Delta / ln(2) \approx 0.42 ~K$ for $L=4$), a reasonable extrapolation of $\Delta(L) / ln(2)$ to $L \to \infty $ gives a correction only $7 \%$ bigger than that for $L = 5$. Interestingly, we can estimate the limiting value of $T_c(L)$ in the presence of dipolar interactions (which we call $T_k^D$ in this limit) by combining this extrapolated value of $\Delta/ln(2)$ obtained from the numerical enthalpy and our mean field relation given by Eq.~\ref{TkDip}. It is this estimated value the one we have used for the scaling of $T_c(L)$ in Fig.~\ref{figure2}(c). The good agreement with the expected behaviour (and with that of the NNSIM) lends further confidence in our procedure and in our molecular field analysis. The value of the temperature shift for a macroscopic system is $\Delta T_k = -0.46 \pm 0.01 ~K$.

\subsection{Phase formation in the low field and temperature regime (DSIM).}

\begin{figure}[htb!]
\centering
 \includegraphics[scale=0.3]{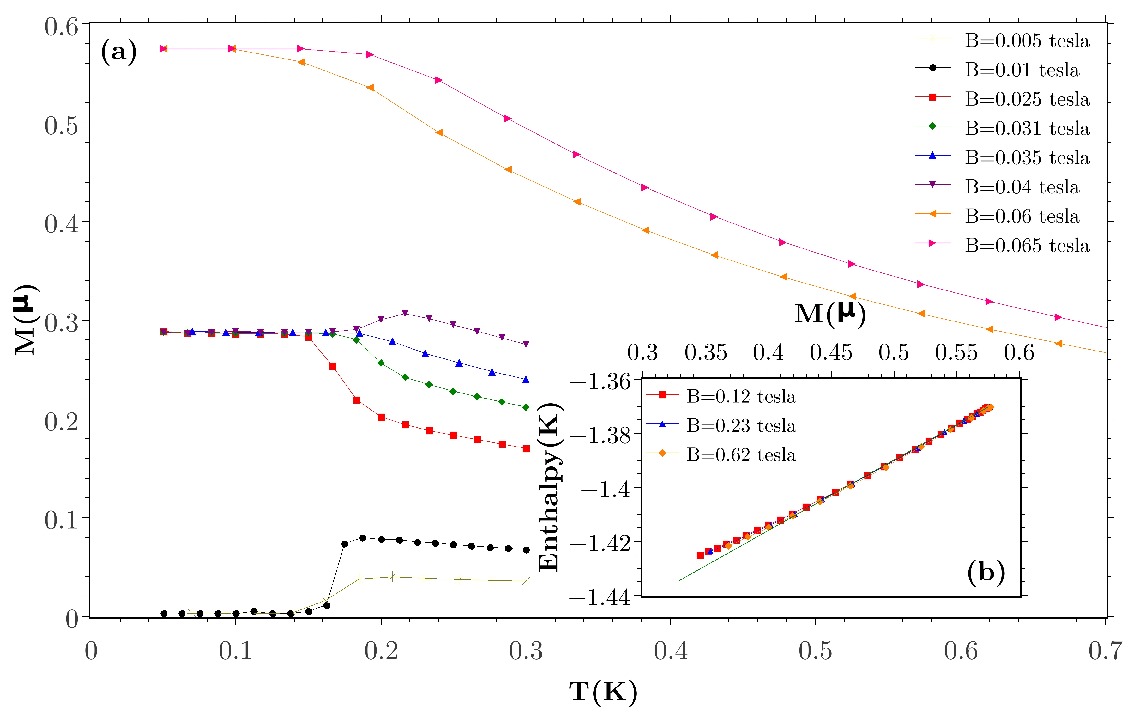}
 \caption{\label{figure4} (a) Magnetisation as a function of temperature for the dipolar model, $L=3$. We observe three plateaus: the first one, at $\approx 0.577 \mu$ corresponds to the saturation of the magnetisation (the FP state), the second plateau at $\approx 0.289 \mu$ is associated with the HP state; the third one at $0$ magnetisation is related to the MDG phase. (b) Enthalpy $U + M B$ for different fields as a function of magnetisation for the dipolar model and lattice size $L=5$. The green line corresponds to the linear fitting near saturation. Its slope represents the enthalpy gain per unit magnetisation associated with the introduction of individual strings of reversed spins on the saturated state. The nonzero value of this slope is entirely due to residual dipolar forces (i.e., those that cannot be assimilated in an interaction between two magnetic charges). }
\end{figure}

We now turn our attention to the subject of phase formation at low temperatures and fields. Firstly, we show that our CMA reproduces well previous results obtained using a more sophisticated algorithm. We then inquire about the possibility of stabilizing other phases not yet found. The main panel of Fig.~\ref{figure4} shows the magnetisation as a function of temperature for different applied fields and $L = 3$, obtained using the CMA. As in previous results (ref.~\cite{taiwan}) we distinguish three plateaux below $\approx$ 0.2 K: i) At high fields, the behaviour is the expected one, with curves reaching the saturation magnetisation value ($\frac{1}{\sqrt{3}}\mu$) after a Kasteleyn transition modified by the dipolar interactions; ii) At very low fields, the curves flatten at $M$=0, in association with the MDG state first found for the DSIM at zero field ~\cite{melko}; iii) At intermediate fields a plateau reaches a magnetisation of half of the saturation value; it corresponds to the HP phase, with a magnetic unit cell represented in Fig.~\ref{semipol}. The existence of two non-saturated ground states in the presence of a field is another obvious consequence of the residual dipolar interactions: when the residual effective field (of order $\Delta B \approx 0.05$ tesla ) becomes stronger than $B$, the Kasteleyn transition cannot take place.


\begin{figure}[htb!]
\centering
 \includegraphics[scale=0.35, angle=0]{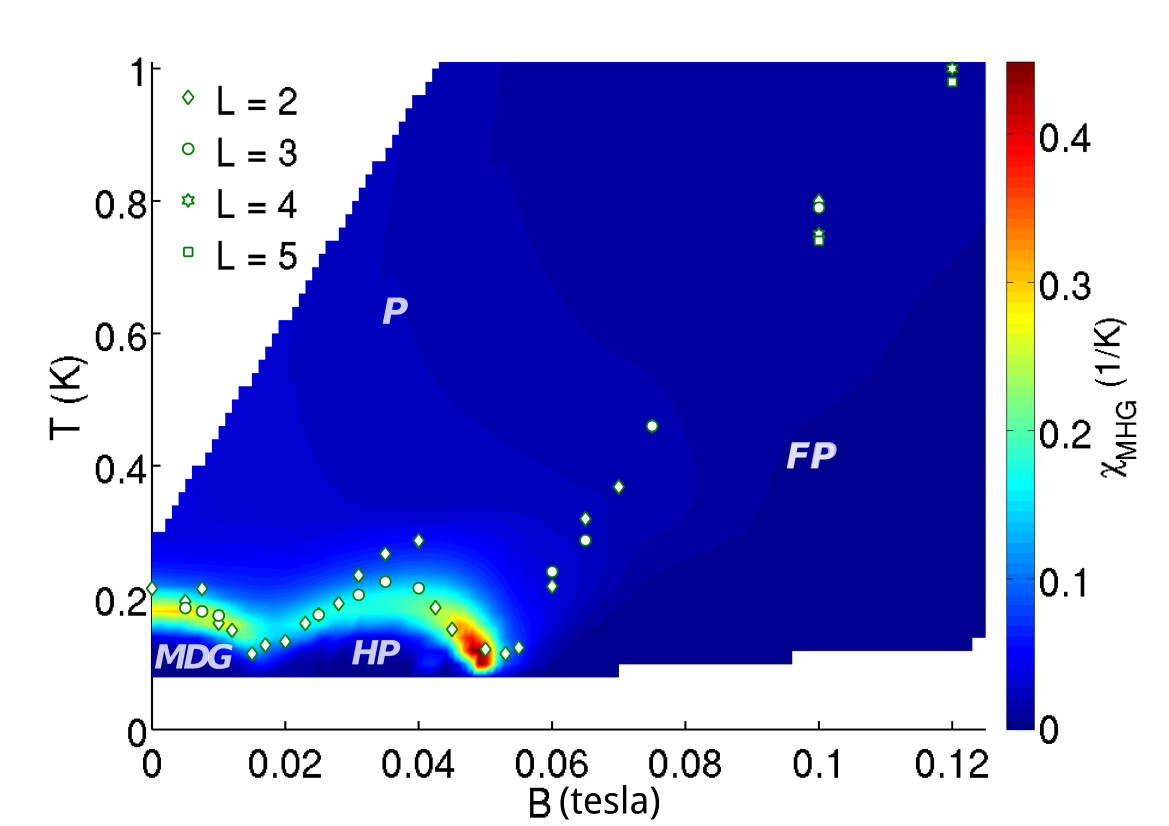}
 \caption{\label{figure6}Phase diagram obtained for the low temperature and low field regime of the dipolar model. We see as an interpolated colour plot the fluctuations on the MDG order parameter \cite{melko} for $L=2$, and indicated in symbols the maxima on the magnetic susceptibility for $L=2, 3, 4,$ and $5$. At low fields and $T$ we observe two arches that correspond to the zero magnetisation MDG and the half polarised (HP) phase. At higher fields, symbols tracing a straight line correspond to the modified Kasteleyn transition, which separates the fully polarised (FP) from the paramagnetic (P) spin ice phase.}
\end{figure}

Fig.~\ref{figure6} is our own version of the phase diagram in the low temperature regime, which summarizes data collected both as a function of field and temperature for different system sizes. The colour plot corresponds to an interpolation of 60 curves for the fluctuation of the order parameter $\eta$ associated with the MDG phase ~\cite{melko}, with $L = 2$. This parameter is defined as:

\begin{equation}
\boldsymbol{\eta}_{\alpha}^m=\frac{1}{N}\left|\sum_{j=1}^{N/4}\sum_{a=1}^{4}\sigma_j^a e^{i\phi_a^m+i\boldsymbol{q}_{\alpha}·\boldsymbol{R}_j}\right|
\label{psi}
\end{equation}

\noindent where $\boldsymbol{q}_{\alpha}$ is a wavevector parallel to one of the cubic axis directions (noted by  $\alpha$), and $m$ labels each of the four possible configurations for each  $\boldsymbol{q}_{\alpha}$ with phase factors $\{ \phi_a^m\}$ ~\cite{melko}. Finally $\boldsymbol{R}_j$ is the position of each up tetrahedron, and $a$ sums over the pseudospins $\sigma_j^a$ in such tetrahedra. 

While this fluctuation does not provide information on the Kasteleyn transition, it does account for the other two phase changes: two concatenated domes can be clearly seen at low temperatures. In order to make the phase diagram more complete, we include symbols corresponding to the maxima on the magnetic susceptibility for different system sizes. These points tend to retrace the previous domes, while they also show the Kasteleyn transition we studied in the previous subsection. Our phase diagram is quite similar to that provided in ref.~\cite{taiwan}.

\begin{figure}[h!]
\centering
 \includegraphics[scale=0.4]{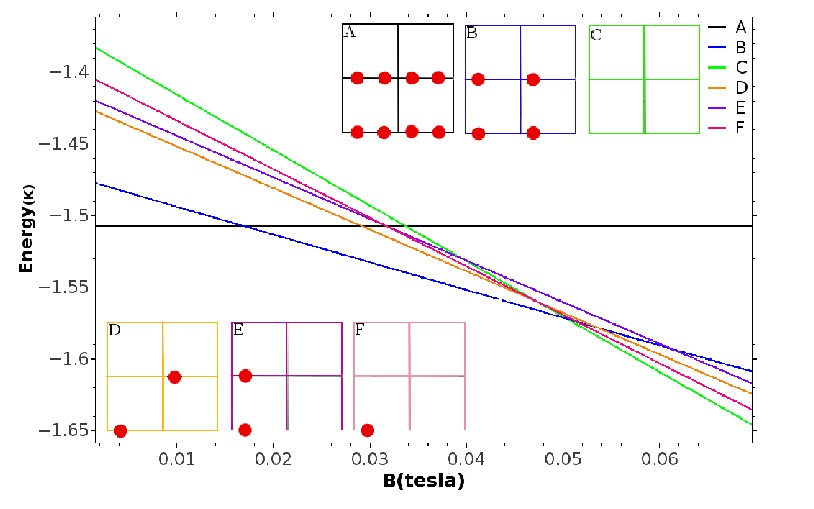}
\caption{\emph{Main figure}: Energy as a function of the intensity of the external magnetic field for different proposed ``string ordered" configurations. The colour of each curve indicates the configuration, a cartoon picture of which is drawn at the top right or bottom left.  \emph{Top right}: the three known ground states represented as ordered string phases (four unit cells are drawn). As in Fig. 1, a straight string of spins reversed from full polarization is represented by a red dot. Configuration \emph{A} corresponds to the MDG state, \emph{B} to the HP phase, and \emph{C} to the FP phase. \emph{Bottom left}: other proposed ground states. As implied by the main panel, the lowest configuration energy always correspond to one of the three known ground states (see \cite{taiwan}). }
 \label{rectas}
\end{figure}

Employing the notation used for the magnetic unit cell in Fig.~\ref{semipol}, we  have included in Fig.~\ref{rectas} (top right) three sketches corresponding to the spin configuration seen for each ordered phase (MDG ($A$), HP ($B$) and FP ($C$)). As before, a red dot indicates a string of reversed spins as seen from the direction of the applied field. We have now drawn four neighbouring unit cells for reasons that will become clear in the next paragraph. Following these schemes it is possible to describe the ground state evolution as a function of decreasing field using just the number of strings of reversed spins per unit cell. Quite importantly, these strings are now \emph{straight} ---running antiparallel to the applied field--- with no associated entropy. Starting from a fully saturated state at high fields (see Fig. \ref{figure6}) the field originated in residual dipolar interactions stabilises at moderate applied fields a single string per unit cell (the HP phase) and finally two of these strings, when the Zeeman energy is negligible compared with the residual dipolar field energy (the MDG phase).

A valid question to formulate is whether or not there are other stable phases with different configurations of straight strings besides the MDG, HP, and FP ones. These other phases might have escaped previous detection in Monte Carlo runs due to their much bigger unit cells. As an example, one might ask: is there a range of magnetic fields where the stable phase has a magnetisation intermediate between the HP and FP phases (for example option \emph{E} in Fig.~\ref{rectas}, bottom left)? Why \emph{half} polarisation is special, and not a quarter of the saturation magnetisation? In order to study these possibilities we have analysed the energy of 18 proposed configurations as a function of field. Using the Ewald method, we calculated the energy for $B = 0$ for many different periodic lattices ($L = 20$), obtaining the y-intercept in  Fig.~\ref{rectas} (main panel) while the slope of each straight line is given by the magnetisation of that configuration. For the sake of simplicity, the figure shows just some of the proposed configurations. As we can see, the only energetically stable phases are the ones already known ~\cite{taiwan}. Furthermore, the predicted ranges of stability do not vary much from those seen in Fig.~\ref{figure6}, obtained for much smaller values of $L$.

\section{Conclusions}

We have shown how a very simple method (the conserved monopoles algorithm) can be used to explore a quasi-spin ice manifold, studying situations not explored before \cite{borzi}. They involve the nearest neighbours spin ice model (where, numerically, we recover all the critical exponents associated with the three dimensional Kasteleyn transition) and the dipolar spin ice model. In both cases a magnetic field was applied along $[100]$.

Taking profit of this CMA we studied how the Kasteleyn transition at moderate and high applied fields is modified by dipolar interactions. Given that this transition occurs for nominally zero monopoles, the effect of dipolar interactions should be zero in the monopole approximation. In other words, changes (if any) should be assigned to the effect of residual interactions  --- corrections to the NNSIM that cannot be assimilated as the interactions between two magnetic charges. We observe that, though in this field regime the dependence of the critical temperature $T_k$ vs field is still a straight line, the effect of the dipolar fields is to reduce $T_k(B)$ with respect to the nearest neighbours case by a significant amount ($\approx 0.45 K$, using the parameters for $Dy_2Ti_2O_7$). This in turn corresponds to effective residual fields of $\approx 0.05$ tesla \emph{opposing} the applied field $\bf{B} // [100]$. An effective field calculation, aided by numerical feedback from the simulations, allowed us to account for this values. Previous experimental results support such a critical temperature shift, providing evidence for the big residual dipolar effects we found. 

Regarding the low field behaviour (i.e., fields comparable to the residual dipolar field) our method served us to recover the phase diagram obtained in a previous work \cite{taiwan}. Reasoning in terms of the non-negligible residual field present inspired us to search for new ordered phases ---with bigger unit cells and thus much harder to find in Monte Carlo simulations on small systems--- with magnetisation intermediate between the fully polarised and the unmagnetised MDG state. We found numerically that no new stable ground states should be expected.


\section{Acknowledgments}

The authors wish to thank P. Holdsworth (who first asked us if the CMA could be used to study this problem), M. Gingras, S. A. Grigera, L. Jaubert, and S. Powel for useful discussions. M. L. B. also wishes to thank the Nordic Institute for Theoretical Physics (NORDITA) for providing a fruitful and stimulating environment. The authors acknowledge financial support from ANPCyT through PICT 2013 N$^{\circ}$2004 and PICT 2014 N$^{\circ}$2618, and Consejo Nacional de Investigaciones Cient\'{\i}ficas y T\'ecnicas (CONICET), Universidad Nacional de La Plata (UNLP), and Consejo Interuniversitario Nacional (CIN).

\end{document}